\documentclass[preprint]
{elsarticle}
\usepackage{graphicx}
\usepackage{dcolumn}
\usepackage{bm}
\usepackage{amsmath, amssymb}
\usepackage{xcolor}

\biboptions{numbers,sort,compress}

\begin{document}
 

\title{Solitary states in multiplex neural networks: onset and vulnerability}

\author[1]{Leonhard Sch\"ulen}

\author[1]{David A. Janzen}

\author[1]{Everton S. Medeiros}

\author[1]{Anna Zakharova \corref{cor1}}\ead{anna.zakharova@tu-berlin.de}

\address[1]{ Institut f\"ur Theoretische Physik, Technische Universit\"at Berlin, Hardenbergstr. 36, 10623 Berlin, Germany}

\cortext[cor1]{Corresponding author}

\begin{abstract}
We investigate solitary states in a two-layer multiplex network of FitzHugh-Nagumo neurons in the oscillatory regime. We demonstrate how solitary states can be induced in a multiplex network consisting of two non-identical layers. More specifically, we show that these patterns can be introduced via weak multiplexing into a network that is fully synchronized in isolation. We show that this result is robust under variations of the inter-layer coupling strength and largely independent of the choice of initial conditions. Moreover, we study the vulnerability of solitary states with respect to changes in the inter-layer topology. In more detail, we remove links that connect two solitary nodes of each layer and evaluate the resulting pattern. We find a highly non-trivial dependence of the survivability of the solitary states on topological (position in the network) and dynamical (phase of the oscillation) characteristics.
\end{abstract}


\maketitle

\section{Introduction}\label{sec:intro}

Synchronization phenomena in networks of coupled oscillators are of great importance in many fields of research ranging from physics and chemistry to biology, neuroscience, physiology, ecology, socio-economic systems, computer science and engineering\cite{PIK01,BAL09,NEK15b,BOC18}. In neural systems, synchronization can play a significant and constructive role in learning and in the context of cognition \cite{SIN99}, but is also linked to pathological states such as Parkinson's disease \cite{POP15} or epilepsy \cite{CHO18, GER20}. It is, therefore, particularly important to understand the mechanisms of synchronization of such systems. Moreover, it crucial to investigate transitions from synchronized states to desynchronized regimes and vise versa, as well as complex partial synchronization patterns \cite{POE15,KRI17} occurring during these transitions. Whereas chimera states \cite{ZAK20}, which represent a peculiar type of partial synchronization pattern defined by a spatial coexistence of synchronous and asynchronous behavior in networks of identical oscillators \cite{ABR04}, have been studied extensively in the context of neuronal networks \cite{MAJ18a,ZAK17a,MIK18,SAW19, SAW19a}, other complex partial synchronization patterns and, in particular, solitary states are still poorly understood.

Solitary states emerge in networks of coupled dynamical units, such as maps \cite{SEM18a,RYB18} and oscillators \cite{MAI14a,JAR18,RYB19a,HEL20} and consist of a large cluster of synchronized oscillators and very few nodes (compared to the network size) that are split off from the synchronized cluster and distributed randomly along the network. The term ``solitary'' stems from the Latin "solitarius" meaning "alone" or "lonely". The solitary elements of the network are "alone" in the sense that the vast majority of the nodes they are coupled to show a uniform behavior different from the dynamics of the node itself. We refer to such solutions as "solitary states" whereas we name the oscillators that are split off from the synchronized cluster "solitary nodes". In other words, "solitary state" always refers to the state of the whole network, while "solitary node" denotes oscillators that do not belong to the synchronized group.

Multilayer networks have recently gained attention of researchers from various fields since they offer a better representation of the topology and dynamics of real-world systems \cite{BOC14,KIV14}. Moreover, they open up new possibilities of control
allowing to regulate nonlinear systems by means of the interplay
between dynamics and multiplexing. On of the advantages of the multiplexing control is  
the possibility of inducing the desired state in one of the layers without manipulating its parameters by solely adjusting the parameters of the other layer. A number of challenging problems occurring in the context of multilayer networks are related to the control of synchronization and complex spatio-temporal patterns. For chimera states, for example, the control through multiplexing \cite{GHO16a,GHO18} and through the interplay of time delay and multiplexing \cite{GHO16,GHO19} has been investigated. Recently, the so-called weak multiplexing control has been reported and applied to chimera states \cite{MIK18} and coherence resonance \cite{SEM18}. The distinctive feature of this control scheme is the possibility of achieving the desired state in a certain layer in the presence of weak coupling between the layers (i.e., the coupling between the layers is much smaller than that inside the layers). There arises a question whether weak multiplexing control can also be applied to solitary states.
Previously, solitary states have been investigated in one-layer neural networks \cite{RYB19a} including those with time-delayed connections \cite{SCH19a}.

In this study, we investigate the onset and possible vulnerabilities of solitary states occurring in a two-layer multiplex network of coupled FitzHugh-Nagumo (FHN) oscillators with non-identical layers. We select the internal control parameters in a way that only one of them is allowed to exhibit solitary states when in isolation, whereas in the other one only the fully synchronized solution is stable in isolation. By letting the two layers interact via a diffusive coupling scheme, we establish the onset of solitary states in the two-layer system for both controlled and random initial conditions. Next, we analyze the robustness of such patterns by investigating their existence for different sets of intra- and inter-layer control parameters. As the solitary states occurring in the two-layer configuration rely on the diffusive interaction between the individual layers, the removal of connections between them (inter-layer links) might pose a threat to solitary states. Indeed, we show that solitary states are vulnerable with respect to the removal of inter-layer links. In more detail, the following two factors play important role: (i) the spatial location of the nodes for which the inter-layer links are removed; (ii) the dynamical phases of the FHN oscillators at this location.

\section{Solitary states in disconnected layers}\label{sec:single}

We start our investigations on the occurrence of solitary states in two-layer networks by considering an isolated layer of FitzHugh-Nagumo (FHN) oscillators non-locally coupled in a ring topology. For such networks, the occurrence of solitary states has been recently discussed, see Ref. \cite{RYB19a}. Therefore, we use this reference as a corner stone for our investigation. The system of equations describing the network of FHN oscillators is given by \cite{RYB19a}:
\begin{equation}
\begin{aligned}
\varepsilon\frac{du_{i}}{dt} &=u_{i}-\frac{u^3_{i}}{3}-v_{i} +\frac{\sigma}{2R}\sum\limits_{j=i-R}^{i+R} [b_{uu}(u_{j}-u_{i})+ b_{uv}(v_{j}-v_{i})] , \\
\frac{dv_{i}}{dt} &= u_{i}+a_i
+ \; \frac{\sigma}{2R}\sum\limits_{j=i-R}^{i+R} [b_{vu}(u_{j}-u_{i})+ b_{vv}(v_{j}-v_{i})],
\end{aligned}
\label{equ:single}
\end{equation}
where $u_{i}$ and $v_{i}$ are, respectively, the activator and inhibitor variables of each FHN oscillator $i$, with $i=1,\dots,N$. The parameter $N$ is the total number of oscillators in the network. The strength of the coupling is given by $\sigma$. The coupling range $R$ indicates the number of nearest neighbors in each direction on the ring. The quantity $R$ can be normalized by the total number of oscillators in the network, allowing us to introduce a quasi-continuous parameter called coupling radius $r=R/N$. For an individual FHN oscillator, the value of the variable $a_i$ defines the excitability threshold and determines whether the system is in the excitable ($|a_i|>1$), or oscillatory ($|a_i|<1$) regime. Since we study networks of identical oscillators, we set $a_i=a=0.5$ (oscillatory) for all $i$. Finally, parameter $\varepsilon$ characterizes time-scale separation between activator $u$ and inhibitor $v$ and is fixed to $\varepsilon=0.05$ throughout the paper.

The coupling function in Eq.~(\ref{equ:single}) contains not only direct, but also cross inputs between activator ($u$) and inhibitor ($v$) variables. This aspect is modeled by a rotational coupling matrix as discussed in \cite{OME13}:
\begin{equation}
B = \left(
\begin{array}{lr}
b_{\mathrm{uu}} & b_{\mathrm{uv}} \\
b_{\mathrm{vu}} & b_{\mathrm{vv}}
\end{array}
\right) =
\left(
\begin{array}{lr}
\cos \phi  & \sin \phi \\
-\sin \phi  & \cos \phi
\end{array}
\right),
\label{equ:Matrix_B}
\end{equation}
where $\phi\in[-\pi;\pi)$. Originally developed in the context of chimera states, it has been shown that solitary states may arise for a variety of values of $\phi$, depending on the coupling strength of the network \cite{RYB19a}. Here we fix the parameter $\phi=\pi/2-0.2$, as this guarantees the occurrence of solitary states for a rather large interval of the coupling strength.

The spatio-temporal patterns occurring in the network described by Eq. (\ref{equ:single}) also depend on the choice of the system's initial conditions (ICs). Specifically, for prescribed values of $\sigma=0.3$ and $\phi=\pi/2-0.2$, the number of solitary nodes in the asymptotic solutions varies with the system ICs. Therefore, we first identify the ICs (basins of attraction) corresponding to the patterns with different number of solitary nodes and then let the system ICs evolve into the desired pattern. This procedure is implemented  by randomly selecting $1$, $5$, $9$ and $13$ oscillators on the ring and assigning to them the initial conditions $u_{sol}(t=0)=1.48421$ and $v_{sol}(0)=0.113235$. The oscillators are chosen in such a way that the cases with less perturbed oscillators are subsets of the cases with more oscillators. For example, the case with $13$ oscillators contains the same oscillators as the one with $9$ plus $4$ additional oscillators. All other oscillators have initial conditions $u_{sync}(0)=-0.501745$ and $v_{sync}(0)=-0.806115$. Naturally, these IC values are not chosen by chance. Instead, we first simulate the system given by Eq. (\ref{equ:single}) for randomly chosen initial conditions and then use the mean of the activator and inhibitor values of the synchronized and the solitary cluster in the last time step as our initial conditions. 

\begin{figure*}
	\centering
	\includegraphics[width=\columnwidth]{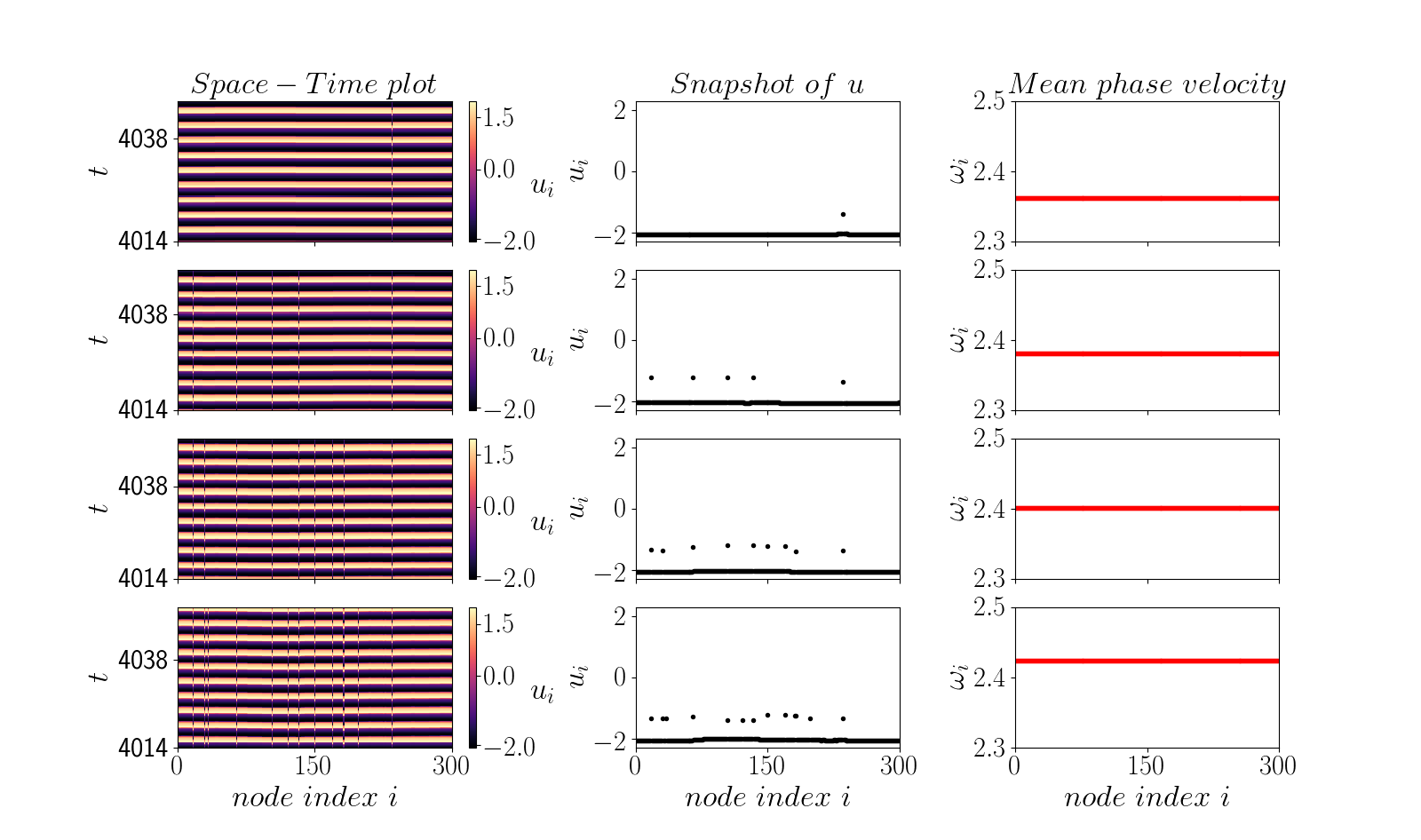}
	\caption{Examples for solitary states in a single layer network of nonlocally coupled FHN units with one (first row), five (second row), nine (third row) and thirteen (fourth row) solitary nodes. The left panels show the space-time plots, the middle ones illustrate snapshots of the activator variable $u_i$ and the right ones depict the mean-phase velocity profiles. Parameters are: $N=300$, $\sigma=0.3$, $\phi=\pi/2-0.2$, $r=0.35$, $a=0.5$, $\varepsilon=0.05$.}
	\label{fig:Fig_1}
\end{figure*}

The resulting spatio-temporal patterns are shown in Fig. \ref{fig:Fig_1} for solutions with $4$ different sizes of solitary cluster. In the left-hand column, we show the space-time evolution of the obtained patterns. The dynamical behavior of the solitary states is visualized in contrast with the synchronous spatial phase. In the middle-column, we show the values of the activator variables at a chosen time (snapshot). One can see that, despite being rather close to the larger cluster, the solitary nodes indeed behave independently from the synchronized ones. The oscillation pattern is persistent in time, as can be seen from the space-time plots. In the right-hand column, we show the mean phase velocity profile for the dynamical behavior of every oscillator in the network. One can immediately notice that the oscillators are, in all cases, frequency synchronized.

\begin{figure}
	\includegraphics[width=\columnwidth]{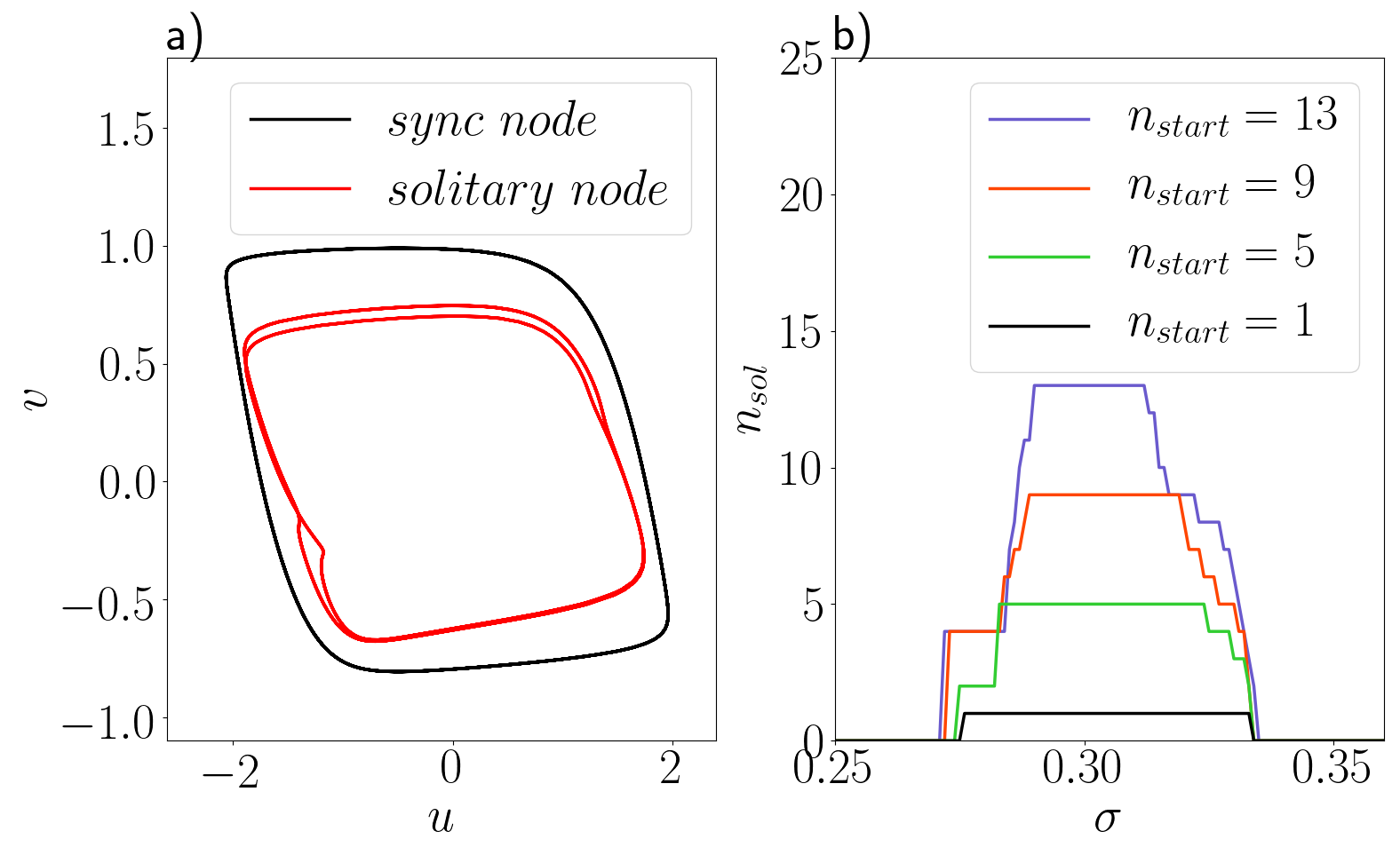}
	\caption{a) Exemplary phase portrait of a solitary state showing the trajectory followed by the synchronized cluster (black) and the one followed by the solitary node (red) with $\sigma=0.3$. b) Number of solitary nodes depending on the coupling strength $\sigma$ for 1 (black), 5 (green), 9 (red) and 13 (blue) initial solitary nodes corresponding to the ones depicted in Fig. \ref{fig:Fig_1}. Other parameters are: $N=300$, $\phi=\pi/2-0.2$,$r=0.35$, $a=0.5$, $\varepsilon=0.05$}
	\label{fig:Fig_2}
\end{figure}

In Fig. \ref{fig:Fig_2}(a), for the solitary state with one single solitary node (see the first row of Fig. \ref{fig:Fig_1}), we show a $2$-dimensional projection of the systems state space in the direction of the solitary node (red) and in the synchronization manifold (black). From this phase portrait we can see that the solitary node follows a smaller limit cycle, a phenomenon that has been shown in Refs. \cite{RYB19a, SCH19a}. Hence, solitary states in our system consist of two clusters, where one is much larger than the other, but having the same frequency.

To understand the impact of the coupling strength $\sigma$ on the onset and termination of solitary states for the four cases shown in Fig. \ref{fig:Fig_1}, we study the number of solitary nodes as function of $\sigma$ in the interval $[0.25,0.36]$. Specifically, in Fig. \ref{fig:Fig_2}(b), we start at $\sigma=0.3$ and increase the strength by $\Delta \sigma = 0.001$ up to $\sigma=0.36$. Similarly, we perform the same procedure in the downward direction by starting again at $\sigma=0.3$ and decreasing the strength until it reaches $\sigma=0.25$. We observe that solitary states in general occur for a limited intervals of $\sigma$. Also, depending on the number of solitary nodes present in the spatio-temporal pattern, the respective interval of $\sigma$ is different. Namely, the plateau sizes for the curves in the case of $13$ (blue), $9$ (red), $5$ (green) and $1$ (black) solitary nodes indicate that the higher the number of solitary nodes in a spatio-temporal pattern, the smaller the respective region of occurrence. These results suggest that the spatio-temporal patterns with lower number of solitary nodes are structurally more persistent than the patterns with large solitary cluster. 

\section{Onset of solitary states in a two-layer network}\label{sec:multilayer}

We now analyze the onset of solitary states in a two-layer multiplex network. The two-layer multiplex architecture is a subclass of multilayer networks, where the only inter-layer connections are between the replica nodes of every layer. Both of the layers consist, therefore, of the same number of elements. The system of equations describing the two-layer multiplex network is given by:
\begin{equation}
\begin{aligned}
\varepsilon\frac{du_{1i}}{dt}&=u_{1i}-\frac{u^3_{1i}}{3}-v_{1i} +\frac{\sigma_1}{2R_1}\sum\limits_{j=i-R_1}^{i+R_1} [b_{uu}(u_{1j}-u_{1i}) + \;b_{uv}(v_{1j}-v_{1i})]  \\
 &+\sigma_{12}(u_{2i}-u_{1i} ), \\
\frac{dv_{1i}}{dt}&=u_{1i}+a  + \frac{\sigma_1}{2R_1}\sum\limits_{j=i-R_1}^{i+R_1} [b_{vu}(u_{1j}-u_{1i}) + \; b_{vv}(v_{1j}-v_{1i})], \\
\varepsilon\frac{du_{2i}}{dt}&=u_{2i}-\frac{u^3_{2i}}{3}-v_{2i} +\frac{\sigma_2}{2R_2}\sum\limits_{j=i-R_2}^{i+R_2} [b_{uu}(u_{2j}-u_{2i}) + \; b_{uv}(v_{2j}-v_{2i})] \\
& + \sigma_{12}(u_{1i}-u_{2i} ),\\
\frac{dv_{2i}}{dt}&=u_{2i} + a  +\frac{\sigma_2}{2R_2}\sum\limits_{j=i-R_2}^{i+R_2} [b_{vu}(u_{2j}-u_{2i})+ b_{vv}(v_{2j}-v_{2i})],
\label{equ:multiplex}
\end{aligned}
\end{equation}
where $u_{1i}$ ($u_{2i}$) and $v_{1i}$ ($v_{2i}$) are the activator and inhibitor variables of the FHN oscillators in the first (second) layer, respectively. Both layers have the same number of elements $N$, thus $i=1,\dots,N$. The control parameters $a=0.5$ and $\varepsilon=0.05$ are the same as in Section \ref{sec:single}. The coupling radii of both layers are identical and also kept from the previous section, i.e., $r_1=r_2=r=0.35$. The intra-layer dynamics is made distinct by mismatching the respective intra-layer coupling strength specified by $\sigma_1$ and $\sigma_2$. In contrast to the intra-layer coupling scheme described in Section \ref{sec:single}, the inter-layer coupling function is of a simple diffusive type controlled by the inter-layer coupling strength $\sigma_{12}$.

\begin{figure}
	\includegraphics[width=\columnwidth]{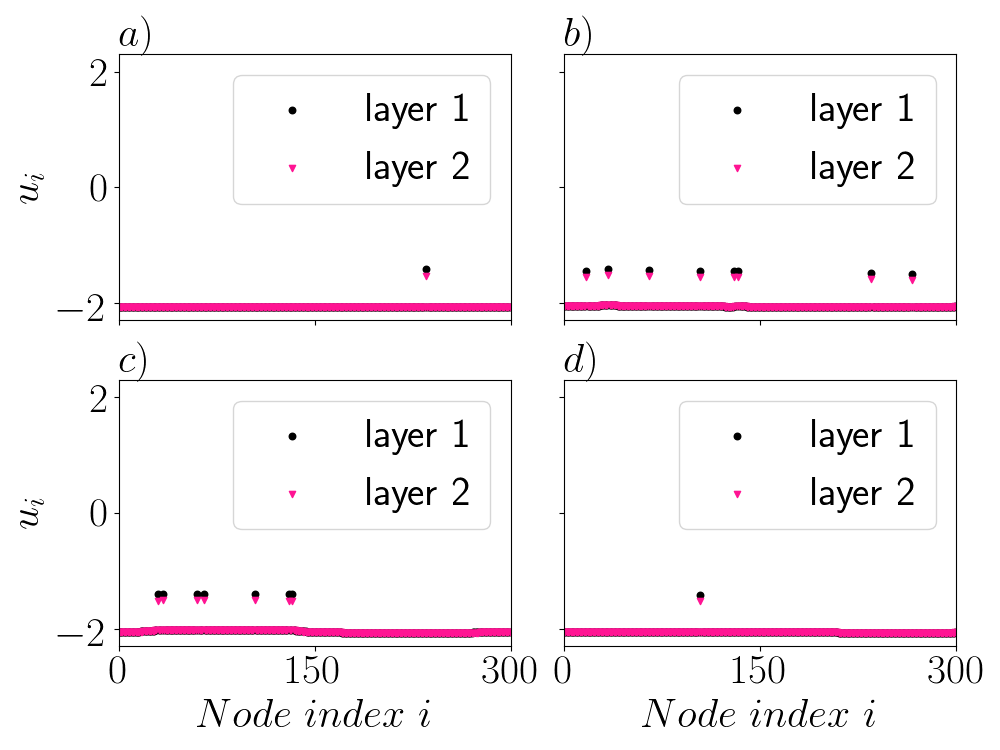}
	\caption{Snapshots of the activator $u_i$ in the first layer (black circles) and second layer (pink diamonds) of solitary states in a two-layer network with the same initial conditions as used in Fig. \ref{fig:Fig_1} in both layers for a) 1, b) 5, c) 9 and d) 13 initial solitary nodes.  Parameters are: $N_1=N_2=300$, $\sigma_1=0.3$, $\sigma_2=0.4$, $\sigma_{12}=0.05$, $\phi=\pi/2-0.2$, $r=0.35$, $a=0.5$, $\varepsilon=0.05$.}
	\label{fig:Fig_3}
\end{figure}

In the following, we investigate the onset of solitary states in the two-layer network (Eq. (\ref{equ:multiplex})) with non-identical layers. First, we fix the intra-layer coupling strength of the second layer to $\sigma_2=0.4$. In accordance to Fig. \ref{fig:Fig_2}(b), regardless of the layer ICs, this value guarantees that this layer is fully synchronized when considered in isolation. Next, in order to observe the generation of solitary states in the coupled layers, we set the coupling strength of the first layer to levels supporting the occurrence of such states. We start with $\sigma_1=0.3$ and inter-layer coupling $\sigma_{12}=0.05$. Then, we proceed to define the ICs of both layers in the same fashion as described in Section \ref{sec:single}, i.e., we prescribe four spatio-temporal patterns with $1$, $5$, $9$ and $13$ solitary nodes. With this, we let the system evolve in time. After discarding a transient phase ($\tau=4000$ arb. time units), in Fig. \ref{fig:Fig_3}, we show snapshots of the obtained pattern for the four different cases a) $1$ b) $5$ c) $9$ and d) $13$ initial solitary nodes. We observe the onset of solitary states in both non-identical layers. The solitary nodes are located at the same positions in both layers for all our simulations. Interestingly, the resulting pattern is different when compared to the first layer in isolation. Most strikingly, the case with $13$ solitary nodes in the isolated first layer collapses to just a state with just one single solitary node in both layers (Fig. \ref{fig:Fig_3}(d)). Further, the pattern with $5$ solitary nodes in the single network develops into the state with $8$ in the multiplex system as shown in Fig. \ref{fig:Fig_3}(b) and the pattern with $9$ solitary nodes in the isolated layer results in a pattern with $7$ in Fig. \ref{fig:Fig_3}(c).

\begin{figure}
	\includegraphics[width=\columnwidth]{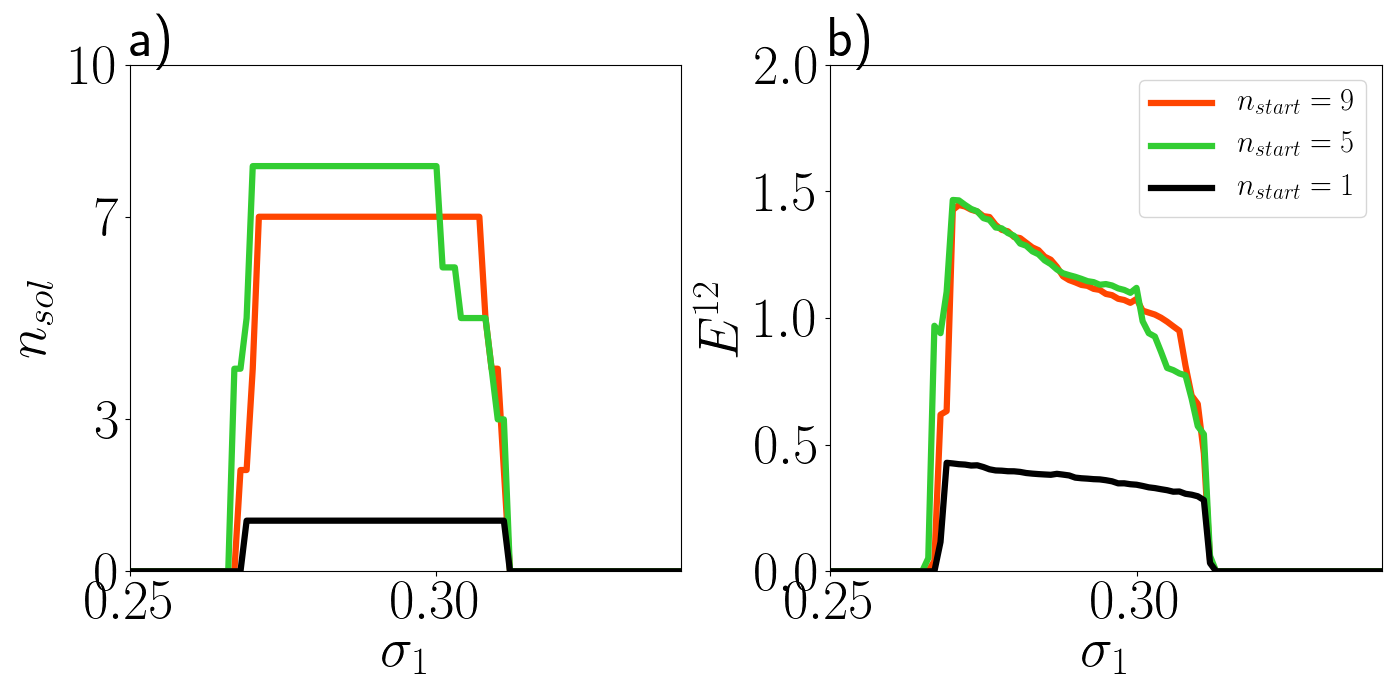}
	\caption{a) Number of solitary nodes depending on the intra-layer coupling strength of the first layer $\sigma_1$ for 1 (black), 5 (green), 9 (red) initial solitary nodes as shown in Fig. \ref{fig:Fig_3}. 
	b) The inter-layer synchronization error $E^{12}$ depending on the intra-layer coupling strength of the first layer $\sigma_1$ for 1 (black), 5 (green), 9 (red) initial solitary nodes as shown in Fig. \ref{fig:Fig_3}. Other parameters are: $N=300$, $\sigma_2=0.4$, $\sigma_{12}=0.05$, $\phi=\pi/2-0.2$,$r=0.35$, $a=0.5$, $\varepsilon=0.05$}
	\label{fig:Fig_4}
\end{figure}

In the next step, we analyze the robustness of the spatio-temporal patterns containing solitary states to variations in the intra-layer coupling strength of the first layer, i.e., the layer inducing such patterns. As already mentioned, we fix the intra-layer coupling strength of the second layer to $\sigma_2=0.4$ and the inter-layer coupling strength to $\sigma_{12}=0.05$. Hence, in Fig. \ref{fig:Fig_4}(a) we show the number of solitary nodes depending on $\sigma_1$. We restrict ourselves to the cases of $1$ (black curve), $5$(green) and $9$ (red) initial solitary nodes. As in Section \ref{sec:single}, we start our investigation with $\sigma_1=0.3$, vary strength by $\Delta \sigma = 0.001$ in the "upward" direction to $\sigma_1=0.34$ and in the "downward" direction until $\sigma_1=0.25$. When compared to the results for this layer in isolation (shown in Fig. \ref{fig:Fig_2}(b)), we find in Fig. \ref{fig:Fig_4}(a) that the lower boundary of the interval of occurrence of solitary states is slightly shifted to the left. The solitary states arise at lower values of $\sigma_1$. For instance, the threshold for the onset of a single solitary node is at $\sigma=0.276$ for an isolated layer, while it is at $\sigma_1=0.269$ for the two-layer network. The slight shift to left is also observed for the other two states. Similarly, the upper boundary of such interval is also left-shifted. However unlike the lower boundary, the threshold here is significantly lower when compared to the first layer in isolation, and surprisingly, all solitary states abruptly collapse at $\sigma_1=0.311$. Nevertheless, similarly to the single layer case, solitary states in the two-layer network exist in a rather narrow region of $\sigma_1$. 

In our approach, the states occurring in the two-layer network are induced by the existence of such states in one of the layers in isolation. The question of how high the level of synchronization between the two heterogeneous layers is, arises therefore naturally, especially given the fact that the solitary nodes are located at the same positions in both layers. In order to address this issue, we estimate the inter-layer synchronization error given by:
\begin{equation}
E^{12} = \lim_{T \to \infty} \frac{1}{NT} \int_0^T \sum_{i=1}^N \|\mathbf{x}^2_i(t) - \mathbf{x}^1_i(t) \| dt,	
\label{equ:error}
\end{equation}
where ${\bf x}_i=(u_{i},v_{i})$, with $i=1,\dots,N$, is the vector containing the activator and inhibitor variables of the FHN oscillators in each layer. The operation $\| . \|$ denotes the euclidean norm. The superscript indices in Eq. (\ref{equ:error}) identify layers. The parameter $T$ indicates the time interval considered in the averaging process. Hence, the definition of $E^{12}$ takes into account differences in the state variables of corresponding oscillators in each layer, lower values of $E^{12}$ indicate higher synchronization level between the layers. In Fig. \ref{fig:Fig_4}(b), we obtain $E^{12}$ for the interval of intra-layer coupling strength $\sigma_1$ at which solitary states occur. We observe that, despite solitary nodes occurring in the same spatial location of both layers, the synchronization among them is imperfect. This fact is expected once the layers are non-identical, namely $\sigma_1\neq\sigma_2$. In accordance, $E^{12}$ decreases as the coupling strength $\sigma_1$ increases towards the value of $\sigma_2$, i.e., the difference between the layers is reduced. On the other hand, the solitary states abruptly cease to exist prior to the identical case $\sigma_1=\sigma_2$.

\begin{figure}
	\includegraphics[width=\columnwidth]{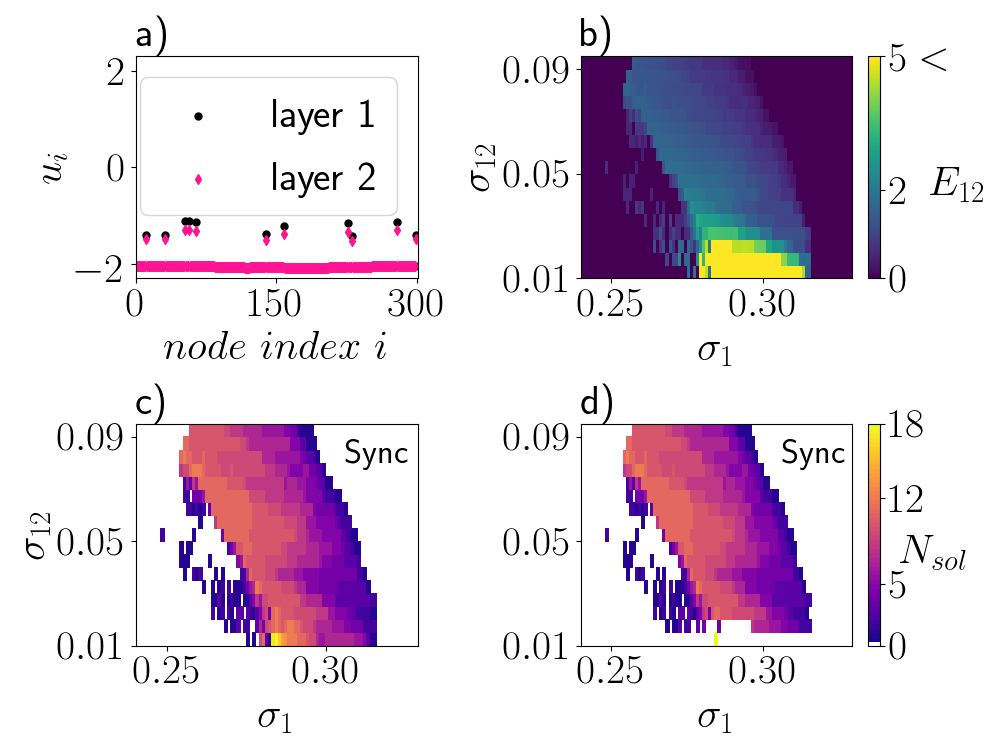}
	\caption{a) Snapshot of the activator $u_i$ of the first layer (black circles) and the second layer (pink diamonds) for $11$ solitary nodes obtained from random initial conditions with $\sigma_1=0.275$ and $\sigma_{12}=0.05$. b) Inter-layer synchronization error for solitary states obtained from random initial conditions and various values of $\sigma_1$ and $\sigma_{12}$. Regions of existence of solitary states obtained from random initial conditions for various values of $\sigma_1$ and $\sigma_{12}$ in layer 1 (c)) and layer 2 (d)). Other parameters are: $N_1=N_2=300$, $\sigma_2=0.4$, $\phi=\pi/2-0.2$ , $r=0.35$, $a=0.5$, $\varepsilon=0.05$ }
	\label{fig:Fig_5}
\end{figure}

Up to this point we concentrated our analysis of solitary states in a two-layer network considering restricted sets of initial conditions for both layers. We now generalize this approach by taking the initial conditions randomly in the state space of the two-layer system. Hence, in Fig. \ref{fig:Fig_5}(a) we show a snapshot of a spatio-temporal pattern containing $11$ solitary nodes obtained from random ICs. We observe that, similarly to the case with specially prepared ICs, the solitary nodes are also at the same position in both layers. This pattern is obtained for $\sigma_1=0.275$ and $\sigma_{12}=0.05$. 

Now, without the necessity of remaining in the basins of attraction of a given pattern as the system parameter is varied, we can to investigate the influence of more than one system parameter on the occurrence of solitary states. Therefore, we analyze the two-layer system in a two-dimensional parameter diagram with $x$-axis being the intra-layer coupling strength of the first layer $\sigma_1$ and the $y$-axis being the inter-layer coupling strength $\sigma_{12}$. Hence, in Figs. \ref{fig:Fig_5}(c) and \ref{fig:Fig_5}(d), we show the occurrence of solitary states obtained from random initial conditions in both the first c) and the second d) layer. On the $x$-axis the intra-layer coupling strength of the first layer $\sigma_1$ is varied in the interval $[0.26, 0.33]$, while on the $y$-axis the inter-layer coupling strength $\sigma_{12}$ is varied in the interval $[0.01, 0.1]$. Color coded is the number of solitary nodes. The white region depicts the completely synchronized state, at which system behaves like a single FHN oscillator. From these figures, we find that for stronger inter-layer coupling strength, the region of existence of solitary states shifts significantly to the left along the $\sigma_1$ axis. This result is in agreement with the comparison between the single network (isolated layer) shown in Fig. \ref{fig:Fig_2}(b) and the two-layer network with $\sigma_{12}=0.05$ shown in Fig. \ref{fig:Fig_4}(a). Another interesting conclusion obtained from Figs. \ref{fig:Fig_5}(c) and Fig. \ref{fig:Fig_5}(d) is that solitary states are observed for any value of $\sigma_{12}$. Only for very small values of $\sigma_{12}$, the interaction between the layers is not strong enough to induce solitary states in the second layer. Furthermore, in the lower left region of these figures, one can see fuzzy behavior typical for multistability. For this parameter region, it is very difficult to ascertain if the system converges to a solitary state or completely synchronizes. Finally, in Fig. \ref{fig:Fig_5}(b), we analyze the quality of the inter-layer synchronization for the studied combinations of $\sigma_1$ and $\sigma_{12}$. The color code represents the error $E^{12}$. We observe that for the parameters corresponding to solitary states, the synchronization error between the layers is higher when compared to the cases at which the layers are internally synchronized. For low values of $\sigma_{12}$, the error $E^{12}$ increases as the pulling between the layers is simply not large enough to ensure their mutual synchronization. Another interesting aspect captured from Fig. \ref{fig:Fig_5}(b) is that a higher number of solitary nodes implies higher values of $E^{12}$.

\section{Phase-sensitivity to inter-layer links removal}

As we demonstrated in the previous sections, the onset of solitary states in the considered two-layer network relies on characteristics of the first layer allowing to induce the solitary states in the second one. One can expect that the systematic removal of inter-layers links would release the second layer and, consequently, the solitary states would eventually remain confined in the first layer. However, despite such a straightforward mechanism, further analysis of such a process reveals a non-trivial dependence of the survivability of solitary states on two characteristics of the system: i) the spatial location of the nodes which inter-layer links are removed; ii) the dynamical phases of the oscillators corresponding to the removed inter-layer link.

Before investigating such dependences, we recall that the spatial location (index $i$) of the nodes containing solitary nodes are identical in both layers. Thus, there are inter-layer links connecting oscillators that are in the synchronized clusters and inter-layer links connecting the solitary nodes of the layers. We refer to the former as "synchronized links" and to the latter as "solitary links". Importantly, we restrict the removal procedure only to solitary links, since those have the strongest impact on the system's spatio-temporal pattern. 

\begin{figure}[!htp]
	\includegraphics[width=\columnwidth]{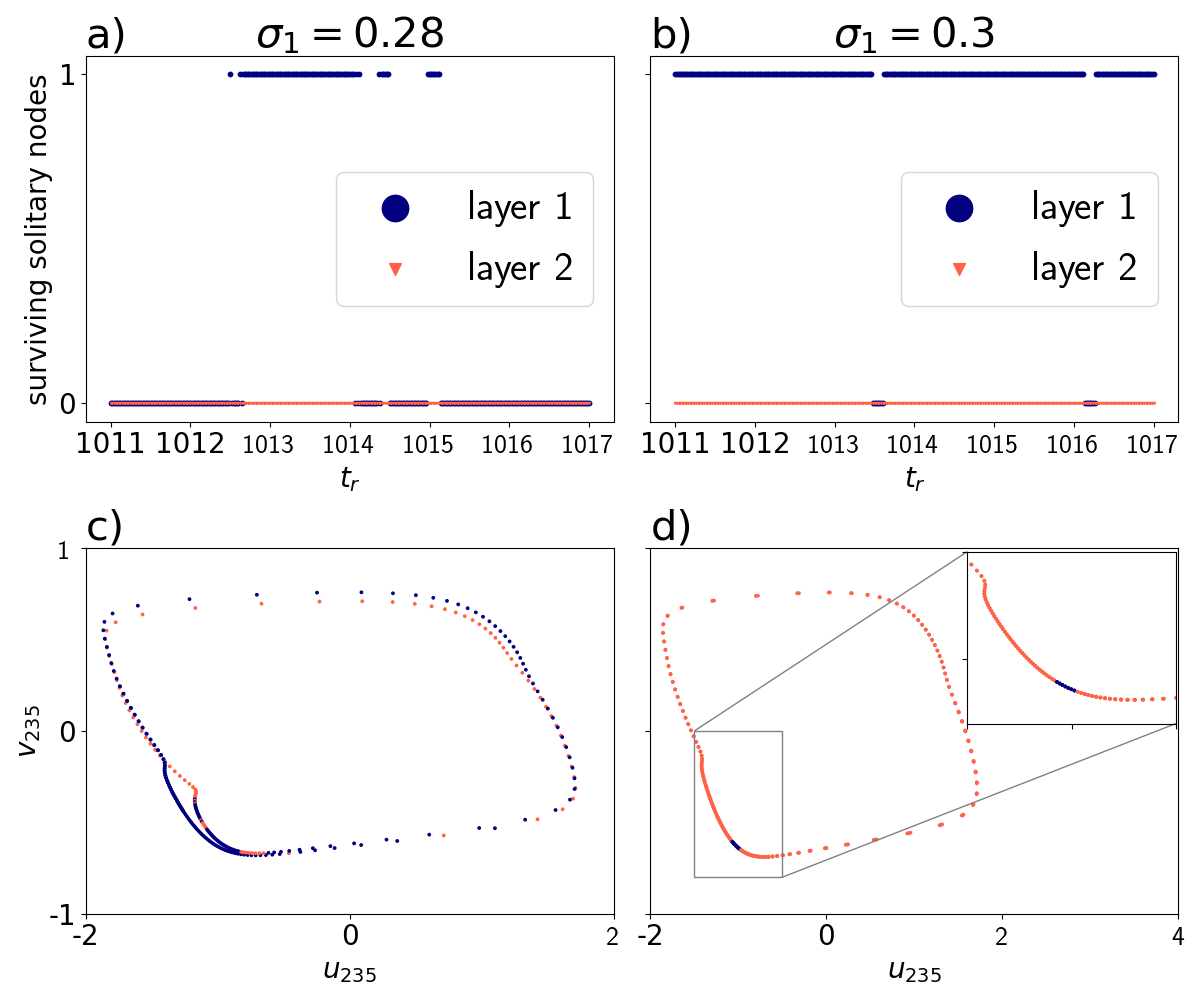}
	\caption{Two cases of vulnerability of a single solitary node for a two-layer multiplex network with a), c) intra-layer coupling strength of the first layer $\sigma_1 = 0.28$ and b), d) $\sigma_1 = 0.3$; a) and b) show the number of surviving solitary nodes for the first (blue circles) and the second layer (red triangle) for different removal times $t_r$.  c) and d) show the dependency of the survival in layer 1 on the position in the phase space where either one (blue) or no (red) node survives as solitary. Other parameters are: $N_1=N_2=N=300$, $\sigma_2 = 0.4$, $\sigma_{12}=0.05$, $a = 0.5$, $\varepsilon = 0.05$, $r_1 = r_2 = 0.35$ , $\phi = \pi/2 -0.2$}

	\label{fig:Fig_6}
\end{figure}

First, we consider the two-layer network containing only one solitary node as shown in Fig. \ref{fig:Fig_3}(a). For this case, we have only one choice for the removal of an inter-layer link. The spatial location plays no role here, as all states with a single solitary node are identical upon index renaming. Therefore we entirely focus on the influence of the oscillation phase in the solitary node. After discarding an initial transient phase of the trajectory ($\tau=1010$ arb. units), we initiate the removal of the solitary link along the system trajectory for times $t_r$ equally spaced in time by $\Delta t=0.01$. After each removal, we evolve the system for another time interval of $\tau$ to check the survivability of solitary states.
Hence, in Fig. \ref{fig:Fig_6}, we investigate the phase sensitivity for two situations in which the two-layer network exhibits a single solitary node, namely for $\sigma_1=0.28$ in the left-hand column of this figure and for $\sigma_1=0.30$ in the right-hand column. Specifically, in Figs. \ref{fig:Fig_6}(a) and \ref{fig:Fig_6}(b), we show the number of surviving solitary nodes in each layer ($y$-axis) as function of the removal time $t_r$ during the period of oscillation ($x$-axis). In blue, we illustrate the number of solitary nodes surviving in the first layer, while in orange is the same quantity for the second layer. As can easily be seen, the survivability of solitary nodes is indeed sensitive to the oscillation phase at which the solitary link is removed. For $\sigma_1=0.28$ in Fig. \ref{fig:Fig_6}(a), we observe intervals of removal times with different sizes corresponding to phases at which the solitary state is extinct in both layers, e.g. $t_r \sim [1011.0,1012.8]$. There are also removal times at which the solitary state survives only in the first layer, e.g. $t_r \sim [1013,1014]$. Since the existence of solitary states in the second layer relies on the interaction with the first one, the orange points in Fig. \ref{fig:Fig_6}(a) confirm that this layer indeed can not sustain the solitary state without the input of the solitary link. For $\sigma_1=0.3$ in Fig. \ref{fig:Fig_6}(b), the survivability of solitary states is less sensitive to the oscillation phase. The solitary state is again vanishing in the second layer for all tested phases, while it persists in the first layer for the majority of phases. Finally, in Fig. \ref{fig:Fig_6}(c) for $\sigma_1=0.28$ and in Fig. \ref{fig:Fig_6}(d) for $\sigma_1=0.3$, we show the phase-space projection $u_{235} \times v_{235}$ in the direction of the solitary node in the first layer. In these figures, the color code stands for the number of surviving solitary nodes, namely one (red) and zero (blue). The most important feature to be highlighted is the location of the higher phase-sensitivity in the left lower corner of the limit cycle. We attribute this sensitivity pattern to trajectory disturbances generated by an unstable equilibrium dwelling nearby.

\begin{figure}[!htp]
	\includegraphics[width=\columnwidth]{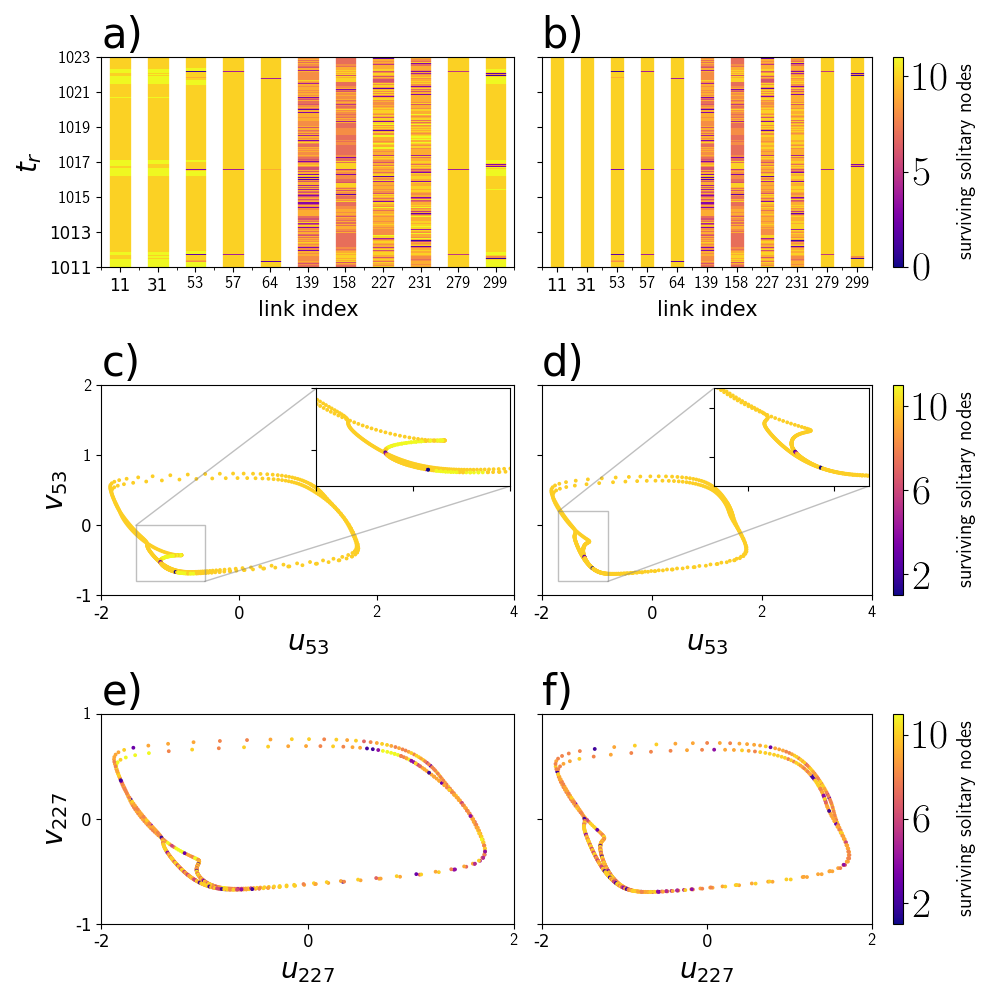}
	\caption{Sensitivity of solitary states depending on the position and the time of removal of the solitary link. a) and b) Color coded number of surviving solitary nodes depending on the position ($x$-axis) and the time ($y$-axis) of the removal of a solitary link. Sensitivity of the number of surviving solitary nodes depending on the phase of the node $53$ (c) and d)) and node $227$ (e) and f)). After the removal of a link, a transient of $\tau=3000$ time units is chosen before evaluating the number of surviving solitary nodes. Other parameters: $N=300$, $\sigma_2 = 0.4$, $a = 0.5$, $\varepsilon = 0.05$, $r_1 = r_2 = 0.35$ , $\phi = \pi/2 -0.2$}
	\label{fig:Fig_7}
\end{figure}

Now, with the insights gathered analyzing the spatio-temporal pattern with only one solitary node, we move on to a more complex pattern containing $11$ solitary nodes as is shown in Fig. (\ref{fig:Fig_5} a). The coupling parameters are $\sigma_1=0.275$ and $\sigma_{12}=0.05$. For this configuration, the dependence of solitary states on the spatial location of solitary links can also be demonstrated. Hence, in order to investigate this dependency, we apply the same removal methodology as in the previous case and we restrict ourselves to the removal of a single solitary link at every network realization. With this, in Figs. \ref{fig:Fig_7}(a) and \ref{fig:Fig_7}(b), we show space-time plots at which the effect of removing different solitary links is visualized for different removal times in the first and second layer, respectively. In these figures, the $x$-axis depicts the indices of the $11$ solitary nodes, while the $y$-axis shows the removal time $t_r$ of the respective solitary link. Color coded is the number of surviving solitary nodes. Note that each time instance and each bar in these plots corresponds to a different network realization. Hence, these plots do not show any kind of time-series, but encapsulate the information for many different simulations. The most important feature of the results shown in Figs. \ref{fig:Fig_7}(a) and \ref{fig:Fig_7}(b) is that the survivability of solitary states depends on a topological (node index) and and a dynamical (phase) characteristics. For instance, comparing the nodes $53$ and $227$ of both layers, we can clearly see that the former shows a regular well-behaved phase-sensitivity, while in the latter the phase-sensitivity is very high. On one hand, for node $53$ the phase at which the link is removed does not matter all that much, the number of surviving solitary nodes is almost always the same, except for phases that are close to the unstable equilibrium. This result is demonstrated in Figs. \ref{fig:Fig_7}(c) and \ref{fig:Fig_7}(d) for the first and second layer, respectively. On the other hand, for node $227$, the picture is drastically different. Here, we observe an extremely high phase-sensitivity, as depicted in Figs. \ref{fig:Fig_7}(e) and \ref{fig:Fig_7}(f). A large variety of phases along with its oscillatory pattern affect the survivability of solitary states.

We attribute the asymmetries observed in the vulnerability of solitary nodes essentially to two different mechanisms: First, the non-uniform distribution of solitary nodes across the spatial extension of our layers. Due to the non-locality of the intra-layer coupling scheme, the removal of a given solitary link indirectly affects a neighborhood of the corresponding node which, in turn, correlates with a different number of solitary nodes. Second, the occurrence of unstable chaotic sets in the systems high-dimensional phase-space results in transients for which the duration depends non-trivially on the spatial direction of perturbations (index of the removed solitary link) and the phase of the respective oscillations at which such perturbations are applied. The occurrence of chaotic transients with different lengths has been previously found to produce intricate phase-sensitivities in synchronized solutions \cite{MED18,MED19}.

\section{Conclusion}

In summary, we report the onset of solitary states for a two-layer network composed of FitzHugh-Nagumo oscillators. For different parameter sets, we found that the number of solitary nodes contained in the solution patterns strongly depends on the choice of the system's initial conditions. Additionally, we have analyzed the structural robustness of the patterns containing solitary nodes by tracking their occurrence for different combinations of the intra- and inter-layer coupling strength. We found that the solitary patterns are persistent for a large continuous region in the two-dimensional parameter diagram composed of these two coupling strengths.

Moreover, we have investigated the vulnerability of solitary states with respect to the removal of inter-layer links. We found that the existence of these states relies on two aspects of the inter-layer links, namely: the spatial location of the nodes connected by for which the inter-layer links are removed and the phase of the oscillations. Specifically, depending on the node location, the removal of the corresponding inter-layer link exhibits different levels of sensitivity to the oscillation phase. Such phase-sensitivity varies from regular patterns, at which the links removal impacts always the same number of surviving solitary node in specific phases, to complex patterns where the number of surviving solitary nodes is unpredictable for any phase along with the oscillation.

\section*{Acknowledgments}
We thank Eckehard Sch\"oll, Yuri Maistrenko, Cristina Masoller and Rico Berner for fruitful discussions. This work was
supported by the Deutsche Forschungsgemeinschaft (DFG, German Research Foundation) - Projektnummer - 163436311 - SFB 910.

\bibliography{ref.bib}
\bibliographystyle{ieeetr}
 
\end{document}